\documentclass[preprint,12pt]{elsarticle}

\usepackage{amssymb}

\usepackage{hyperref}

\journal{Solid State Communications}

\begin{document}

\begin{frontmatter}

\title{Magnetization and the Concurrence of the Spin-1/2 Ising-Heisenberg Pyrochlore Ladder}

\author{A. Sadrolashrafi\fnref{label1}, N. S. Ananikian\fnref{label1}, and L. N. Ananikyan}

\address{A. I.  Alikhanyan National Science Laboratory, 0036 Yerevan, Armenia}

\fntext[label1]{A. Sadrolashrafi: afsaneh@mail.yerphi.am}

\fntext[label2]{N. S. Ananikian: ananik@yerphi.am}

\date{\today}

\begin{abstract}
We have established a quantum antiferromagnetic Heisenberg-Ising model on a spin-1/2 pyrochlore edge-shared ladder with Heisenberg intra-rung and Ising inter-rung interactions as a perspicuous candidate to exhibit magnetization mid and zero plateaus, characteristic peaks of magnetic susceptibility, and thermal entanglement mid plateau. The model is exactly solvable and thus, all the essential properties such as the thermal entanglement and the magnetic properties of the system can be exactly calculated. The calculations are done both through the transfer matrix technique and through the reduced density matrix. The magnetization plateaus are observed at zero and half the saturation value and the magnetic susceptibility exhibits a clear demonstration of the associated characteristic peaks. The model also displays the mid plateau of the thermal entanglement as a function of the external magnetic field at low temperatures. 
\end{abstract}

\begin{keyword}
quantum spin model, pyrochlore ladder, magnetization plateaus, thermal entanglement
\end{keyword}

\end{frontmatter}

\section{INTRODUCTION}
The appearance of the magnetization plateau was predicted in the pioneering theoretical work of Hida \cite{Hida} for a ferromagnetic - ferromagnetic - antiferromagnetic Heisenberg chain $ \mathrm{3CuCl_2 2} $ 2 dioxane compound. Magnetic ordering at low temperatures is frustrated by the geometry of the crystalline lattice, a situation known as geometrical magnetic frustration. The magnetization plateau, specific heat, and magnetic susceptibility of low-dimensional quantum spin systems have attracted much attention over the last few decades both experimentally \cite{Experimental Observation1, Azurite, Frustrated Q Magnets Azurite} and theoretically \cite{Frustrated trimer chain model, distorted diamond chain, Geometric frustration diamond chains, Ananikian azurite, Ananikian diamond chain01, Ananikian diamond chain1, Ananikian diamond chain2, Ananikian diamond chain3, Ananikian diamond chain4}. The frustrating properties of the corner-sharing or edge-shared tetrahedron lattice have been particularly studied for the magnetic pyrochlore oxides \cite{Mag pyrochlore oxides, Stuffed rare earth pyrochlore solid solutions, Mn6 cluster exhibiting, Jaubert}. 

In condensed matter, magnetic materials are of particular interest. Among these we will study the antiferromagnetism through description by Heisenberg models. The pyrochlore edge-shared tetrahedron ladder with spin-1/2 is an excellent candidate to realize the antiferromagnetic properties \cite{Magnetization plateaux and jumps in a class of frustrated ladders} and test the reasonableness and validity of the theories on the quantum entanglement in frustrated systems. The ideas have been taken from the recent article \cite{three-leg tube, triangular tubes} by exactly solving the problem through the classical transfer matrix method and figuring out the spin frustration and thermal entanglement of the spin-1/2 Ising-Heisenberg three-leg tube, which accounts for the Heisenberg intra-triangle and Ising inter-triangle interaction.

We have applied the separation of Heisenberg intra-rung and Ising inter-rung interactions on a pyrochlore ladder with antiferromagnetic spin-1/2 couplings using the classical transfer matrix method and constructed the thermal entanglement, magnetization plateau and magnetic susceptibility as on a diamond chain \cite{Ananikian diamond chain5, Ananikian diamond chain6, Ananikian diamond chain7, Ananikian diamond chain8} . The entanglement properties, the correlation functions, and magnetic properties are also studied in \cite{Afsaneh1, Afsaneh2, Afsaneh3} for a spin zigzag ladder with the generalized Majumdar-Ghosh model, a spin ladder, and a Heisenberg spin chain correspondingly. The functional dependence of the entanglement on the correlation functions and magnetic susceptibility were discussed in Ref.'s \cite{Magnetic susceptibility - entanglement witness, Entanglement and correlation, Dynamics of entanglement in one-dimensional spin systems, Quantum entanglement in Nitrosyl iron complexes, Exp determination of thermal ent in spin clusters mag suscep, trimer spin-1/2 Heisenberg chains, entanglement in bulk properties of solids, Thermal ent in the nanotubular system, Exp observ of ent in low dimensional spin sys, Exp detection of therm ent in a molecular chain}. 

This paper is organized as the following: in the next section we present the antiferromagnetic spin-1/2 Ising-Heisenberg with frustrated magnetization plateau on a pyrochlore edge-shared ladder. Further, in the third section, we have discussed the thermal concurrence, correlation functions and magnetic susceptibility.
\section{Pyrochlore edge-shared ladder: The Model and Its Exact Solution}
\begin{figure}[htbp]
\begin{center}
\includegraphics[scale=0.42]{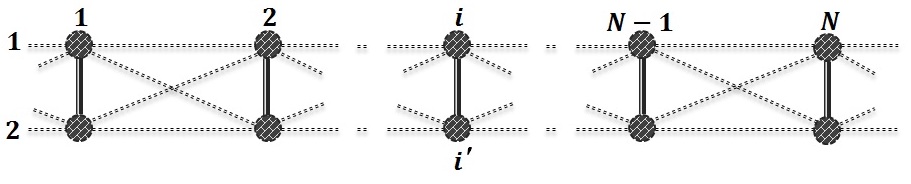}
\caption{The spin-1/2 Ising-Heisenberg model on a pyrochlore edge-shared ladder; Heisenberg intra-rung and Ising inter-rung couplings}
\label{pyrochlore spin ladder}
\end{center}
\end{figure}

\begin{figure}[htbp]
\begin{center}
\includegraphics[scale=0.7]{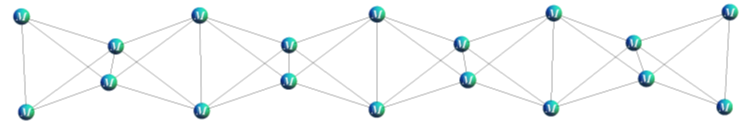}
\caption{tetrahedral structure of the spin-1/2 Ising-Heisenberg model on a pyrochlore edge-shared ladder}
\label{tetra pyro ladder}
\end{center}
\end{figure}

Let us consider the spin-1/2 Ising-Heisenberg model on a ladder, whereas the spins belonging to the same rung are mutually coupled through the Heisenberg intra-rung interaction and the spins from the neighboring rungs are coupled through the Ising inter-rung interaction (see FIG. \ref{pyrochlore spin ladder}).
The Hamiltonian of the spin-1/2 Ising-Heisenberg ladder of $N$ rungs is then given by 
\begin{equation}\label{Tot Ham}
H = \sum_{i=1}^{N}\sum_{\alpha=x, y, z} J_H S_{i}^\alpha S_{i'}^\alpha + J_I \sum_{i=1}^{N} (S_{i}^z + S_{i'}^z) (S_{i+1}^z S_{i+1'}^z) - h \sum_{i=1}^{N} (S_{i}^z + S_{i'}^z),
\end{equation}
where the index $i$ labels the rungs of the ladder with the periodic boundary condition such that the site N+1 would become equal to the first site. The prime-less indices label the spins on the upper leg whereas the primed indices label them on the lower leg; Thus, the spin at site $i$th is the spin on the $i$th rung and on the upper leg whereas the $i'$ show the spin on the same rung but on the lower leg. $J_H$ and $J_I$ are the Heisenberg and Ising coupling constants correspondingly and $h$ is the magnetic field strength. $S_i$ is the spin operator at site $i$ and $S_{i}^\alpha$ is its $\alpha$ component with $\alpha = x, y, z$. Figure (\ref{tetra pyro ladder}) shows the tetrahedron structure of the edge-shared pyrochlore ladder.

The tetrahedron structure of the edge-shared pyrochlore ladder with antiferromagnetic behavior is also noticed in \cite{ueda, koga}.
\subsection{Transfer-Matrix Solution} \label{Transfer-Matrix Solution}
The total Hamiltonian (\ref{Tot Ham}) of the spin-1/2 Ising-Heisenberg ladder can be alternatively rewritten in terms of composite spin operators, which determine the total spin of the Heisenberg rungs and its z-component
\begin{equation}\label{Comp Op}
T_i = S_i + S_i' \ \ \ \ \ , \ \ \ \ T_{i}^{\alpha^2} = \frac{1}{2} + 2 S_{i}^{\alpha} S_{i'}^{\alpha}.
\end{equation}

It can be shown that the composite spin operators $T_{i}^{2}$ and $T_{i}^{z}$ commute with the total Hamiltonian (\ref{Tot Ham}), i.e. $[ H , T_{i}^{ 2} ] = [ H , T_{i}^{z} ] = 0$, which means that the total spin of the rung and its z-component represent conserved quantities with well defined quantum numbers. Consequently, the eigenvalues of the total Hamiltonian (\ref{Tot Ham}) can be related to the eigenvalues of the total spin operator of the rung $T_{i}^{2}$ and its z-component $T_{i}^{z}$.
\\

Using the identity in (\ref{Comp Op}), the total Hamiltonian (\ref{Tot Ham}) can be rewritten in the following form: 
\begin{equation}\label{Tot Ham and Hi}
H = \frac{ - 3 N J_H }{4} + \sum_{i=1}^{N} H_{i},
\end{equation}
in which $H_i$ is the Hamiltonian of the two subsequent rungs at $i$ and $i + 1$ and is written as follows in terms of the total spin operators of the rungs and their corresponding z-components:
\begin{equation}\label{Tot Ham and Hi}
H_i = \frac{J_H}{4} \sum_{i=1}^{N} (T_{i}^2 + T_{i+1}^2) + J_I \sum_{i=1}^{N} T_{i}^{z} T_{i+1}^{z} - \frac{h}{2} \sum_{i=1}^{N} (T_{i}^z + T_{i+1}^z).
\end{equation}

It can be seen that the commutation relation $[ H_i , H_j ] = 0$ is true for any $i$ and $j$ and therefore, $H_i$'s are actually separable from each other. Thus, the spin-1/2 Ising-Heisenberg ladder defined by the Hamiltonian (\ref{Tot Ham}) can be rigorously mapped onto some classical composite spin chain model, which can be further treated by the transfer-matrix method \cite{Baxter} and the relative partition function can be factorized into the following form:
\begin{eqnarray}\label{partition function}
Z &=& tr e^{- \beta H} \nonumber \\
&=& e^{\frac{ 3 N \beta J_H }{4} } \ tr e^{ - \beta \sum_{i=1}^{N} H_{i}} \nonumber \\
&=& e^{\frac{ 3 N \beta J_H }{4} } \ tr \prod_{i=1}^{N} e^{ - \beta H_{i}},
\end{eqnarray}
where $\beta = 1/ ( K_B T )$ shows the inverse of the absolute temperature with $K_B$ being the Boltzman constant. Furthermore, we can consider the following matrix representation for $e^{ - \beta H_i}$ in the basis of the eigenstates of the composite spin operators $T_{i}^{2}$, $T_{i}^{z}$, $T_{i + 1}^{2}$, $T_{i + 1}^{z}$ of the two consecutive rungs, by which we can figure out the transfer matrix $W$ as follows:
\begin{eqnarray}\label{W}
W [ i, i + 1 ]&=&\Big< T_{i}^{2} , T_{i}^{z} | e^{- \beta H_i} | T_{i + 1}^{2} , T_{i + 1}^{z} \Big> \nonumber \\ \nonumber \\
&=& 
\left[ \begin{array}{cccc}
	e^{- \beta ( J_H + J_I + h )} & e^{- \beta ( J_H + h/2 )} & e^{- \beta ( J_H - J_I )} & e^{- \beta ( \frac{J_H + h}{2} )}\\
	e^{- \beta ( J_H + h/2 )} & e^{- \beta J_H} & e^{- \beta ( J_H - h/2 )} & e^{- \beta J_H/2}\\
    	e^{- \beta ( J_H - J_I )} & e^{- \beta ( J_H - h/2 )} & e^{- \beta ( J_H + J_I - h )} & e^{- \beta ( \frac{J_H - h}{2} )}\\
    	e^{- \beta ( \frac{J_H + h}{2} )} & e^{- \beta J_H/2} & e^{- \beta ( \frac{J_H - h}{2} )} & 1\\ 
    \end{array} \right] .
    \end{eqnarray}
    
In the above representation, we chose $ \big< T_{i}^{2} , T_{i}^{z} |$'s and $ \big< T_{i + 1}^{2} , T_{i + 1}^{z} |$'s from the set $\{ \big< 1 , -1 |, \big< 1 , 0 |,\big< 1 , 1 |,\big< 0 , 0 | \}$. As $H_i$'s dependence on $T_i$'s and $T_{i + 1}$'s is independent of the site $i$ and the ladder is translational invariant, one can rewrite the last line in (\ref{partition function}) as
\begin{equation} \nonumber
Z = e^{\frac{ 3 N \beta J_H }{4} } \ tr W^N .
\end{equation}
Thus, the partition function in the thermodynamic limit $N \rightarrow \infty$ can be solely determined by the largest eigenvalue, $\lambda_{Max}$, of the transfer matrix W given by Eq. (\ref{W}) such that
\begin{equation} \nonumber
Z = e^{\frac{ 3 N \beta J_H }{4} } \ \lambda_{Max}^{N} .
\end{equation}
Then the magnetization per site and the magnetic susceptibility can be calculated through the following formulae:
\\
\begin{eqnarray} \nonumber
m&=&- \frac{1}{ N } \frac{ \partial F }{ \partial h } = \frac{1}{ N \beta } \frac{ \partial \log{Z} }{ \partial h } = \frac{1}{ \beta } \ \frac{ \partial \log{ \lambda_{Max} }}{\partial h}\\
\chi&=&\frac{ \partial m }{ \partial h } = \frac{ \partial^2 \log{ \lambda_{Max} } }{\partial h^2} 
\end{eqnarray}

In section \ref{the Magnetization Plateaus in the Antiferromagnetic Couplings}, we will present the corresponding graphs that show the behavior of the magnetization and also the susceptibility of the pyrochlore ladder (\ref{Tot Ham}) with the antiferromagnetic couplings of $J_H = 3/2$ and $J_I = 1$. But before we proceed to the section \ref{the Magnetization Plateaus in  the Antiferromagnetic Couplings}, in \ref{The Reduced Density Matrix of One Rung} we present a second approach to calculate the magnetization $m$ and the susceptibility $\chi$, based on the information derived from the reduced density matrix of the rung. This latter approach would be beneficial when one aims to investigate the relation between the magnetic properties of the system with the quantum correlations and the entanglement.

It appears that for the pyrochlore ladder (\ref{Tot Ham}), the results from the two approaches i.e. the transfer matrix approach and the reduced density matrix approach would fully correspond with one another.
\subsection{The Reduced Density Matrix of One Rung}\label{The Reduced Density Matrix of One Rung} 
For a rung of two spin-1/2 particles in the pyrochlore ladder (\ref{Tot Ham}) the thermal reduced density matrix of which can be derived from the density matrix of a block of two adjacent rungs will have the following form in the standard basis:
\begin{equation}\label{Redrho}
\rho = 
\left(
\begin{array}{cccc}
z & 0 & 0 & 0 \\
 0 & x & y & 0 \\
 0 & y & x & 0 \\
 0 & 0 & 0 & w \\
\end{array}
\right)
\end{equation}
\\
with $x$, $y$, $z$, and $w$ being the following functions of the coupling constants $J_H$ and $J_I$, the external magnetic field $h$, and the inverse absolute temperature $\beta = \frac{1}{ K_B T }$ with $K_B$ being the Boltzman constant:
\\
\begin{eqnarray}\nonumber
x&=& \frac{ e^{ - \frac{ \beta}{2} ( 2 J_H + h ) } }{2} \left( e^{ \frac{ \beta J_H }{2}} + 1 \right) \left( e^{ \frac{ \beta}{2} ( J_H + h )} + e^{ \beta h} + e^{ \beta h/2} + 1 \right)\\ \nonumber
y&=&- \frac{ e^{ - \frac{ \beta}{2} ( 2 J_H + h ) } }{2} \left( e^{ \frac{ \beta J_H }{2}} - 1 \right) \left( e^{ \frac{ \beta}{2} ( J_H + h )} + e^{ \beta h} + e^{ \beta h/2} + 1 \right)\\ \nonumber
\\ \nonumber
z&=&e^{ - \beta ( J_H + J_I )} \left( e^{ \frac{ \beta}{2} ( J_H + 2 J_I + h )} + e^{ \beta ( J_I + \frac{ h }{2} )} + e^{ 2 \beta J_I} + e^{ \beta h} \right)\\ \nonumber
\\ \nonumber
w&=&e^{ - \beta ( J_H + J_I + h )} \left( e^{ \frac{ \beta}{2} ( J_H + 2 J_I + h )} + e^{ \beta ( 2 J_I + h )} + e^{ \beta ( J_I + \frac{ h }{2} )} +1 \right) \nonumber
\end{eqnarray}
\\

As soon as the above reduced density matrix is ready, the rest of the calculation is straightforward: To achieve the magnetization per site, $m$, one needs to calculate the expectation value of the operator $S_z$, $\frac { 1 }{ tr ( \rho ) } tr ( \rho S_z)$, where $S_Z$ is the z-component of the spin operator of one particle at a site. Then the next step would be the derivative with respect to the external magnetic field, $h$, which yields the magnetic susceptibility of the ladder. Thus, one can follow the following formulae to figure out the magnetic properties of the system:
\begin{eqnarray}\nonumber
m&=&\frac { 1 }{ tr ( \rho ) } tr ( \rho S_z)\\
\chi&=&\frac{ \partial m }{ \partial h } = \frac{ \partial }{ \partial h } \ \Big( \frac { 1 }{ tr ( \rho ) } tr ( \rho S_z) \Big)
\end{eqnarray}
\\
\subsection{the Magnetization Plateaus and the Magnetic Susceptibility in the Antiferromagnetic Couplings}\label{the Magnetization Plateaus in the Antiferromagnetic Couplings}
The phenomenon of magnetization plateau has been studied during the past decade both experimentally and theoretically. The plateau may exist in the magnetization curves of quantum spin systems in the case of a strong magnetic external field at low temperatures. 

The phenomenon of magnetization plateau is considered as a macroscopic manifestation of the essentially quantum effect in which the magnetization $m$ is quantized at fractional values of the saturation magnetization $m_s$ in low dimensional magnetism \cite{Hida, ferromagnetic-ferromagnetic-antiferromagnetic spin chain, Haldane Gap for Half-Integer Spins, Breakdown of the magnetization plateau, trimerized XXZ spin chain, polymerized spin chains, Antiferromagnetic Zigzag Spin Chain in the Strongly Frustrated Region, Fractional Sz excitation in Ising-like zigzag XXZ chain, anisotropic s=1/2 antiferromagnetic chain}. 

The quantum plateau state was actually first discovered, over two decades ago in the ferromagnetic-ferromagnetic-antiferromagetic (F-F-AF) chain model \cite{Hida, ferromagnetic-ferromagnetic-antiferromagnetic spin chain, Breakdown of the magnetization plateau, trimerized XXZ spin chain}. 

Magnetization plateaus appear in a wide range of models on chains, ladders, hierarchical lattices and theoretically analyzed by dynamical, transfer matrix approaches as well as by the exact diagonalization in clusters \cite{solid 3He system on a triangular lattice, ZIGZAG LADDER, zigzag ladder2, ferromagnetic-ferromagnetic-antiferromagnetic Ising chain, Mag and Lyapun on a kagome chain}. 

To explain the experimental measurements of magnetization plateau and the double peak behavior in the natural mineral azurite, there have been proposed different types of theoretical Heisenberg models (the density-matrix and transfer-matrix renormalization- group techniques, density functional theory, high-temperature expansion, variation mean-field-like treatment, based on the Gibbs-Bogoliubov inequality) \cite{Frustrated trimer chain model, spin-1/2 Heisenberg diamond chains, dimer-monomer, Ananikian azurite}. 

Magnetization plateaus and the multiple peak structure of the specific heat have also been observed on an Ising-Hubbard diamond chain \cite{Ananikian diamond chain01}.

In the previous parts we introduced two separate approaches to achieve the magnetic properties of the system; One is based on the transfer matrix method, which is explained in \ref{Transfer-Matrix Solution} and the other one is based on the calculations upon the quantum density matrix, which is talked about in \ref{The Reduced Density Matrix of One Rung}. As it was mentioned earlier, the two approaches yield exactly the same results for our spin-1/2 pyrochlore ladder. 

\begin{figure}[htbp]
\begin{center}
\includegraphics[scale=0.2]{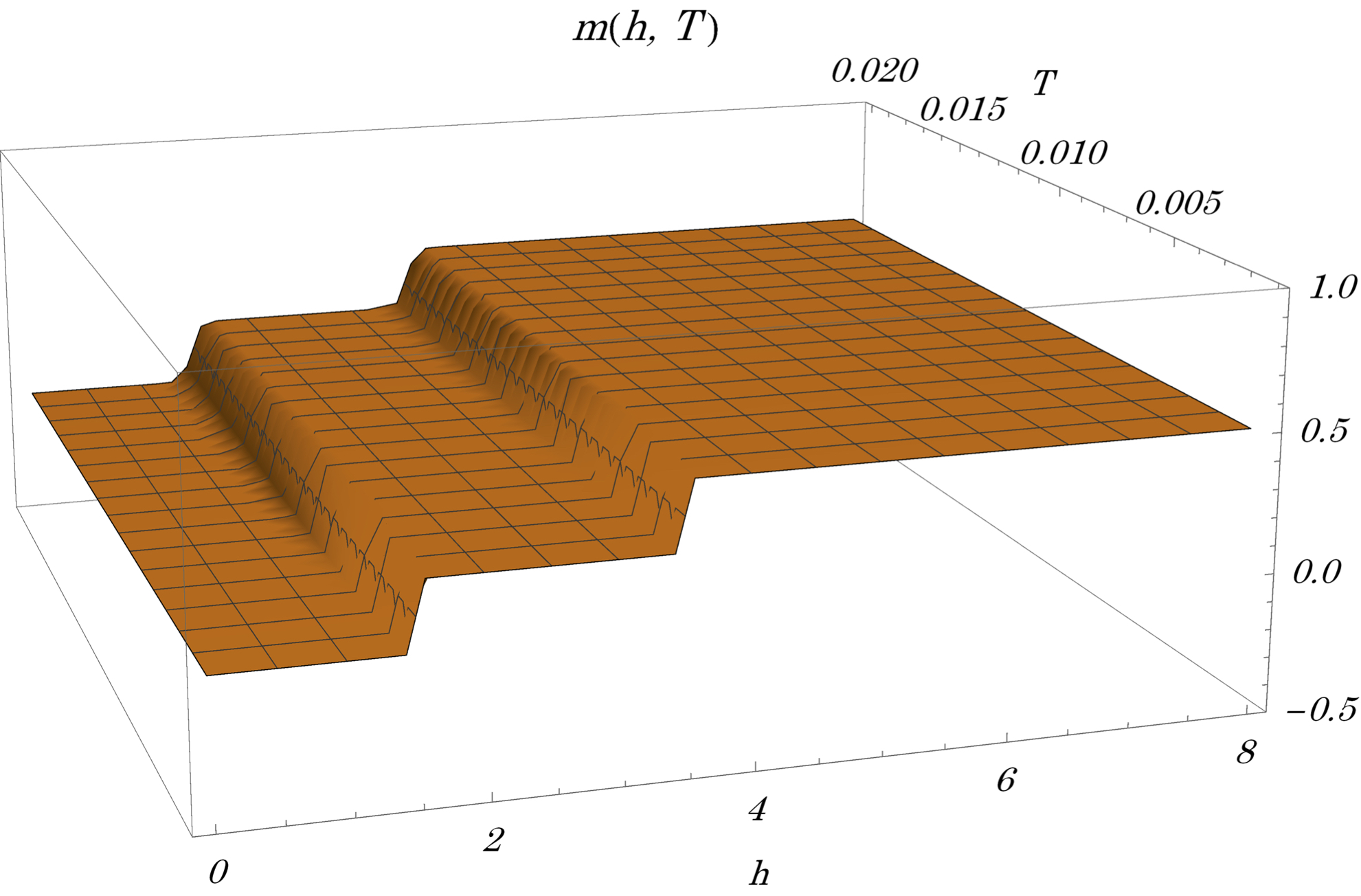}
\caption{the magnetization behavior of the spin-1/2 Ising-Heisenberg pyrochlore edge-shared ladder with Heisenberg coupling constant $J_H = 3/2$ and Ising coupling constant $J_I = 1$, as a function of the absolute temperature $T$ and the magnetic field $h$; The Boltzman constant $K_B$ has been set to 1. The existence of the magnetization plateaus are clear in the low temperatures.}
\label{m}
\end{center}
\end{figure}

\begin{figure}[htbp]
\begin{center}
\includegraphics[scale=0.2]{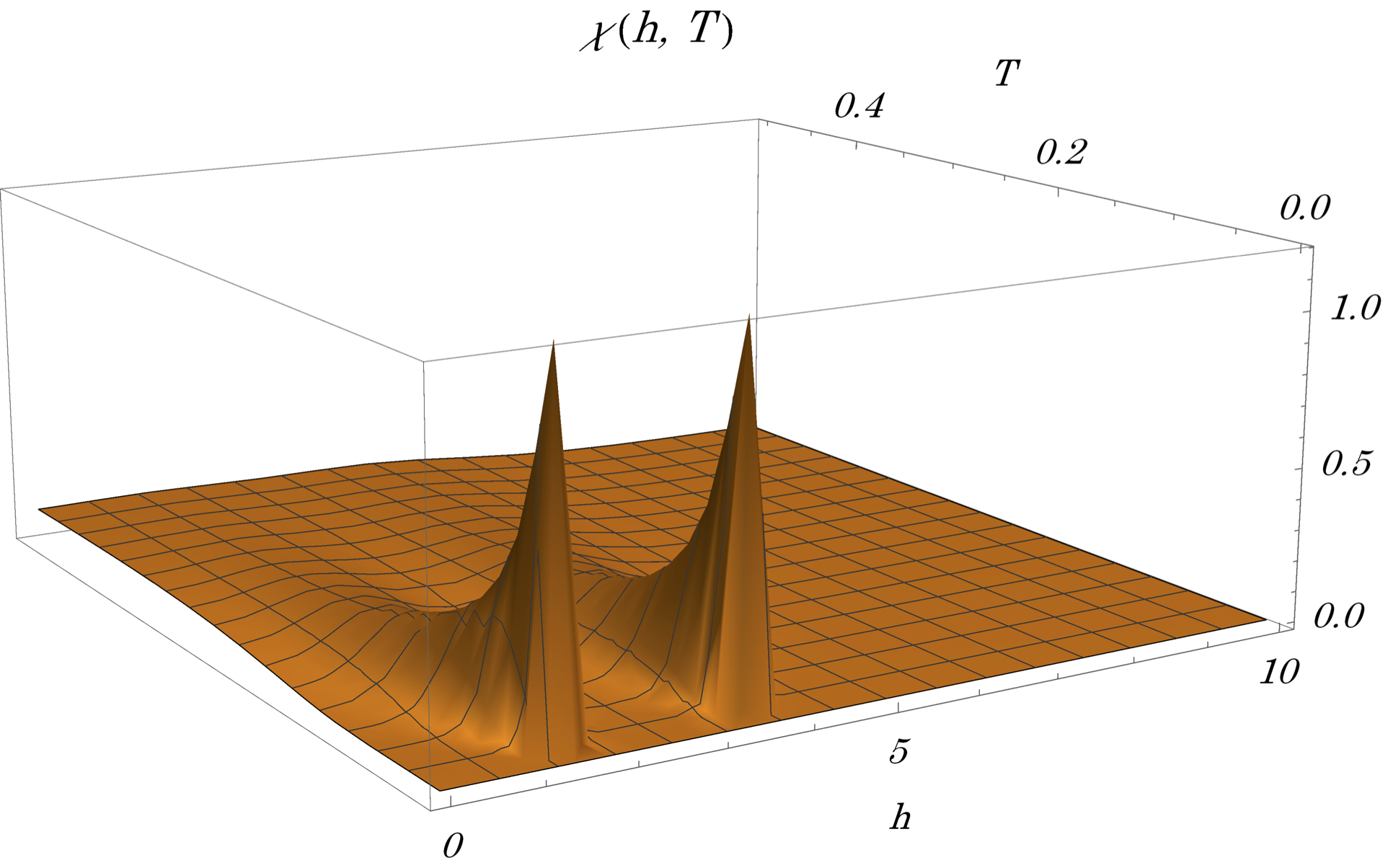}
\caption{Magnetic susceptibility of the spin-1/2 Ising-Heisenberg pyrochlore edge-shared ladder with Heisenberg coupling constant $J_H = 3/2$ and Ising coupling constant $J_I = 1$, as a function of the absolute temperature $T$ and the magnetic field $h$; The Boltzman constant $K_B$ has been set to 1. The characteristic peaks of the magnetic susceptibility can be seen in the low temperatures.}
\label{suscep}
\end{center}
\end{figure}

Figures (\ref{m}) and (\ref{suscep}) show these results for the magnetization per site and the magnetic susceptibility of the ladder with the Heisenberg interaction and Ising interaction coupling constants being $J_H = 3/2$ and $J_I = 1$ correspondingly. The existence of the magnetization plateaus at zero and half of the saturation magnetization at low temperatures is clear in Fig.(\ref{m}) and the corresponding characteristic peaks of the magnetic susceptibility are shown in the Fig.(\ref{suscep}). 

\section{Entanglement of the Antiferromagnetic Pyrochlore Ladder}\label{Entanglement of the Antiferromagnetic Pyrochlore Ladder}
Entanglement is a type of correlation that is quantum mechanical in nature. It reflects nonlocal correlations between particles, even if they are removed and do not directly interact with each other. 

Studying entanglement in condensed matter systems is of great interest due to the fact that some behaviors of such systems can most probably only be explained with the aid of entanglement. The magnetic susceptibility at low temperatures, quantum phase transitions, chemical reactions are examples where the entanglement is the key ingredient for a complete understanding of the system. Furthermore, in order to produce a quantum processor, the entanglement in condensed matter systems becomes an essential concept.

In order to measure the entanglement between two spin-half particles sitting on the same rung in the Ising-Heisenberg pyrochlore ladder, we study the concurrence of the two Heisenberg qubits, using the definition proposed by Wootters et al. \cite{Wootters, Hill}. 

The concurrence can be expressed in terms of a matrix $R$ in the following manner:
\begin{equation}\nonumber
R=\rho\cdot\left(\sigma^{y}\otimes\sigma^{y}\right)\cdot\rho^{*}\cdot\left(\sigma^{y}\otimes\sigma^{y}\right),
\end{equation}
which is constructed as a function of the density operator $\rho$, given by (\ref{Redrho}), with $\rho^{*}$ being the complex conjugate of $\rho$ and $\sigma^{y}$ being the Pauli $Y$ matrix $\left(\begin{array}{cccc} 0 & - i \\ i & 0 \\ \end{array} \right)$.

Thereafter, the concurrence of two Heisenberg qubits (the bipartite entanglement) can be obtained in terms of the eigenvalues of the Hermitian positive matrix $R$:
\begin{equation} \nonumber
\mathcal{C}(\rho) = \mathrm{max} \ \{ \sqrt{\lambda_{1}} - \sqrt{\lambda_{2}} - \sqrt{\lambda_{3}} - \sqrt{\lambda_{4}} \ , 0 \}
\end{equation}
with $\lambda_{1} \geqslant \lambda_{2} \geqslant \lambda_{3} \geqslant \lambda_{4}$.
\\

\begin{figure}[htbp]
\begin{center}
\includegraphics[scale=0.2]{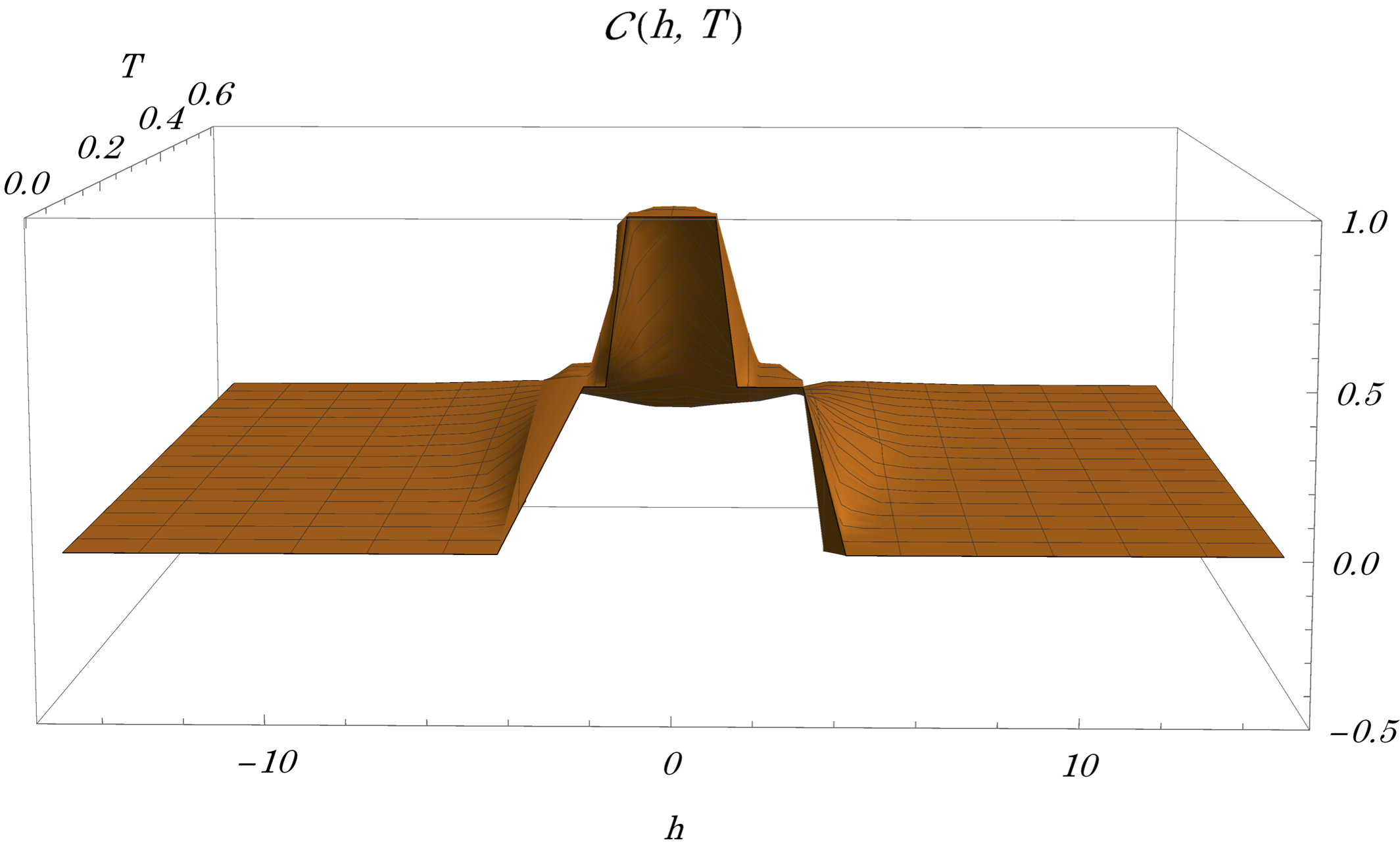}
\caption{the concurrence between the two sites on a rung in the ladder (\ref{Tot Ham}) with Heisenberg coupling constant $J_H = 3/2$ and Ising coupling constant $J_I = 1$ }
\label{concurr}
\end{center}
\end{figure}

\begin{figure}[htbp]
\begin{center}
\includegraphics[scale=0.2]{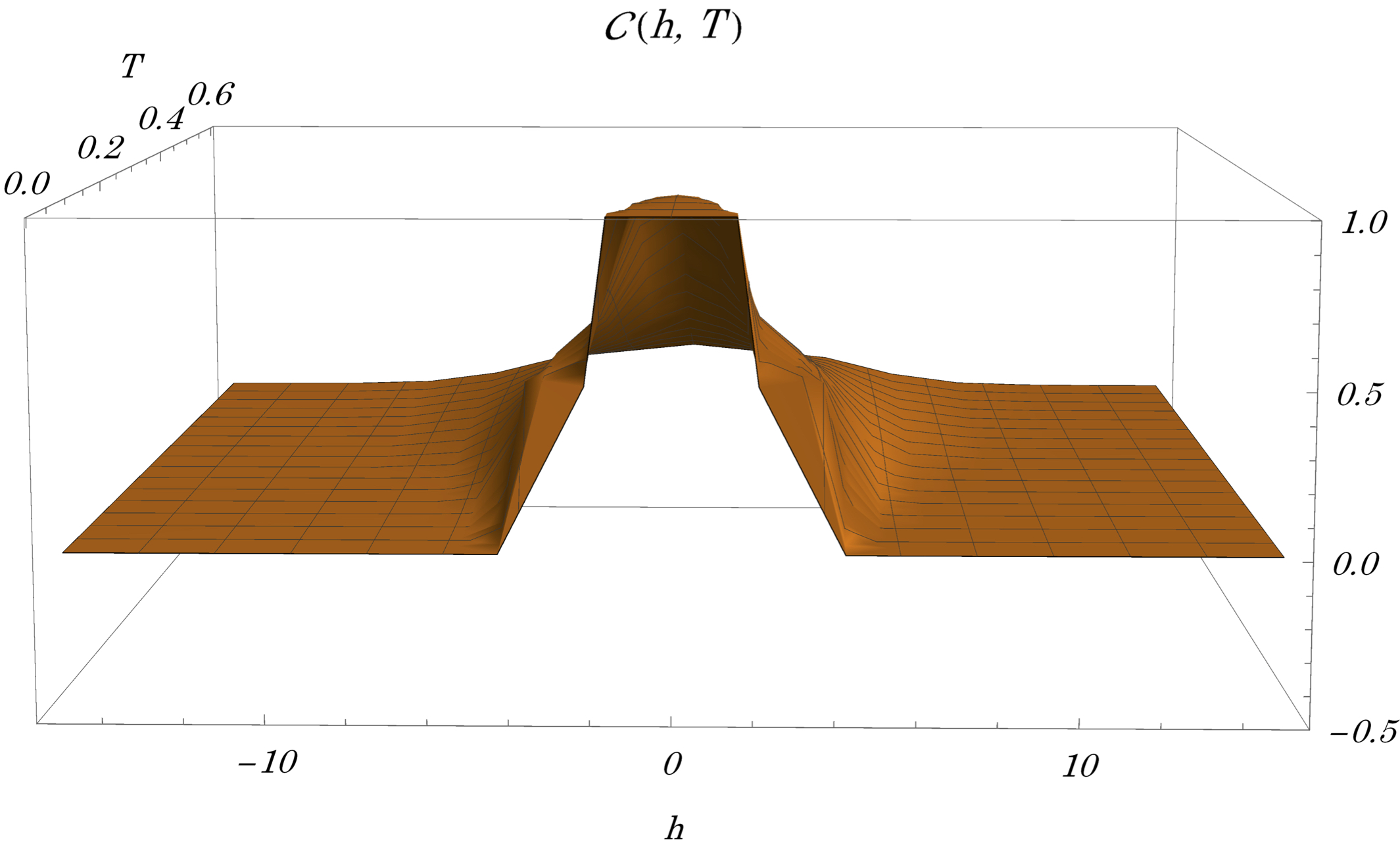}
\caption{the concurrence between the two sites on a rung in the ladder (\ref{Tot Ham}) with Heisenberg coupling constant $J_H = 2$ and Ising coupling constant $J_I = 1$}
\label{concurr2}
\end{center}
\end{figure}

Accordingly, the rung concurrence is shown in figures (\ref{concurr}) and (\ref{concurr2}) as a function of the absolute temperature $T$ and the external magnetic field $h$ with the Heisenberg and Ising coupling constants being correspondingly 3/2 and 1 in Fig. (\ref{concurr}) and 2 and 1 in Fig. (\ref{concurr2}). In both cases, the appearance of mid plateaus at low temperatures are evident.

M. Wie\'{s}niak, V. Vedral, and C. Bruckner showed in 2005 how to relate entanglement with the magnetic susceptibility \cite{ Magnetic susceptibility - entanglement witness}. In 2008, exploiting the Blaney and Bowers equation \cite{paramagnetism of copper acetate}, S. M. Aldashin studied the entanglement in dimer systems \cite{Quantum entanglement in Nitrosyl iron complexes} and gained the results which relate the entanglement with the magnetic susceptibility of the system. The entanglement of a dimer-trimer system was studied experimentally using magnetic susceptibility in 2008 by M. Souza et al. \cite{Exp determination of thermal ent in spin clusters mag suscep}. In \cite{trimer spin-1/2 Heisenberg chains}, in 2015, entanglement for the trimer compound is shown to be related to the magnetic susceptibility of the material. 

On the other hand, it is a difficult task to determinate experimentally if a state is entangled or not.  A widely used method for that purpose, for entanglement detection is the use of an Entanglement Witness (EW)\cite{Horodecki, Horodecki2}. An observable $W$ can be used as an EW if $tr (\rho W) < 0$ when $\rho$ is an entangled state and if $tr (\rho W) \geqslant 0$ then $\rho$ may or may not be entangled. Magnetic susceptibility was proposed as an EW \cite{ Magnetic susceptibility - entanglement witness}, and several experimental results were obtained within this framework \cite{Exp determination of thermal ent in spin clusters mag suscep, entanglement in bulk properties of solids, Thermal ent in the nanotubular system, Exp observ of ent in low dimensional spin sys, Exp detection of therm ent in a molecular chain}. 

As it is rather straightforward to calculate the magnetic susceptibility and thermal entanglement over the separable blocks, we believe our proposed model of a pyrochlore edge-shared ladder can be a good candidate to be investigated more in that respect.
\\
\section{Conclusion}
In conclusion, we have established an antiferromagnetic spin-1/2 pyrochlore edge-shared ladder with Heisenberg intra-rung and Ising inter-rung interactions as an appropriate candidate to exhibit magnetization plateaus, characteristic peaks of magnetic susceptibility, and thermal entanglement. The model shows the full block separability and thus, it is capable of being treated as an exactly solvable quantum model, where all the physical quantities including thermal entanglement and the magnetic properties of the system can be exactly calculated. In order to perform the calculations, we exploited two approaches: one is based on the transfer matrix technique and the other is built upon the reduced density matrix of one separable block of two rungs. The magnetization plateaus and the characteristic peaks of the magnetic susceptibility are presented for special values of the interaction constants $J_H = 3/2$ and $J_I = 1$. It is evident that the pyrochlore ladder (\ref{Tot Ham}) displays intermediate magnetization plateaus at zero and also one half of the saturation magnetization. By the calculations upon the reduced density matrix, the thermal entanglement is obtained and its behavior as a function of the absolute temperature $T$ and the external magnetic field $h$ is shown for two sets of the interaction constants, $\{ J_H = 3/2 , J_I = 1 \}$ and $\{ J_H = 2 , J_I = 1 \}$, where the entanglement mid plateaus can be observed at low temperatures.
\section*{Acknowledgments}
The authors acknowledge H. Poghosyan. N. A. acknowledges financial support by the MC-IRSES no. 612707 (DIONICOS) under FP7-PEOPLE-2013 and CS MES RA in the frame of the Research Project no. SCS 15T-1C114 grants. A. S. acknowledges the financial support from the fellowship granted by ICTP Office of External Activities (OEA) at the ICTP affiliated center at Yerevan, Armenia within NET68 and  OEA-AC-100 programs.
\bibliographystyle{elsarticle-harv} 

\begin{thebibliography}{00}
\bibitem{Hida} Kazuo Hida, J. Phys. Soc. Jpn. 63, pp. 2359-2364 (1994)
\bibitem{Experimental Observation1} H. Kikuchi, Y. Fujii, M. Chiba, S. Mitsudo, T. Idehara, T. Tonegawa, K. Okamoto, T. Sakai, T. Kuwai, H. Ohta, Phys. Rev. Lett. 94 (2005) 227201
\bibitem{Azurite} K. C. Rule, A. U. B. Wolter, S. S\"{u}llow, D. A. Tennant, A. Br\"{u}hl, S. K\"{o}hler, B. Wolf, M. Lang, J. Schreuer, Phys. Rev. Lett. 100, 117202 - Published 19 March 2008
\bibitem{Frustrated Q Magnets Azurite} Harald Jeschke, Ingo Opahle, Hem Kandpal, Roser Valentí, Hena Das, Tanusri Saha-Dasgupta, Oleg Janson, Helge Rosner, Andreas Br\"{u}hl, Bernd Wolf, Michael Lang, Johannes Richter, Shijie Hu, Xiaoqun Wang, Robert Peters, Thomas Pruschke, Andreas Honecker, Phys. Rev. Lett. 106, 217201 - Published 23 May 2011
\bibitem{Frustrated trimer chain model} A. Honecker, A. Lauchli, Phys. Rev. B 63 (2001) 174407
\bibitem{distorted diamond chain} H. -J. Mikeska, C. Luckmann, Phys. Rev. B 77 (2008) 054405
\bibitem{Geometric frustration diamond chains} Lucia \u{C}anov\'{a}, Jozef Stre\u{c}ka, Michal Ja\u{s}\u{c}ur Published 5 May 2006, Journal of Physics: Condensed Matter, Volume 18, 4967
\bibitem{Ananikian azurite} N. Ananikian, H. Lazaryan, M. Nalbandyan, Eur. Phys. J. B 85 (2012) 223
\bibitem{Ananikian diamond chain01} M. Nalbandyan, H. Lazaryan, O.Rojas, S. M. de Souza, N. S. Ananikian, J. Phys. Soc. Jpn. 83 (2014) 074001
\bibitem{Ananikian diamond chain1} N. Ananikian, V. Hovhannisyan, Physica A 392 (2013) 2375 
\bibitem{Ananikian diamond chain2} N. S. Ananikian, V. Hovhannisyan, R. Kenna, Physica A 396 (2014) 51
\bibitem{Ananikian diamond chain3} V. Hovhannisyan, N. Ananikian, R. Kenna, Physica A 453 (2016) 116-130
\bibitem{Ananikian diamond chain4} V. Hovhannisyan, J. Stre\u{c}ka, N. Ananikian, J. Phys.: Condens. Matter 28 (2016) 085401 (7pp)
\bibitem{Mag pyrochlore oxides} Jason S. Gardner, Michel J. P. Gingras, a John E. Greedan, Rev. Mod. Phys. 82, 53 - Published 26 January 2010
\bibitem{Stuffed rare earth pyrochlore solid solutions} G. C. Lau, B. D. Muegge, T. M. McQueen, E. L. Duncan, R. J. Cava, J. Sol. St. Chem. 179 (2006), 3126-3135
\bibitem{Mn6 cluster exhibiting} Piya Seth and Ashutosh Ghosh, RSC Adv., 2013, 3, 3717-3725
\bibitem{Jaubert} L. D. C. Jaubert, Owen Benton, Jeffrey G. Rau, J. Oitmaa, R. R. P. Singh, Nic Shannon, Michel J. P. Gingras, Phys. Rev. Lett. 115, 267208 - Published 29 December 2015
\bibitem{Magnetization plateaux and jumps in a class of frustrated ladders} A. Honecker, F. Mila, M. Troyer, Eur. Phys. J. B 15 (2000) 227
\bibitem{three-leg tube} J. Stre\u{c}ka, R. C. Al\'{e}cio, M. L. Lyra, O. Rojas, J. Magn. Magn. Mater. 409, 124 (2016)
\bibitem{triangular tubes} R. C. Al\'{e}cio, M. L. Lyra, J. Stre\u{c}ka, Journal of Magnetism and Magnetic Materials, Volume 417, 1 November 2016, Pages 294-301
\bibitem{Ananikian diamond chain5} Onofre Rojas, M Rojas, N. S. Ananikian, S. M. de Souza, Phys. Rev. A 86  (2012) 042330
\bibitem{Ananikian diamond chain6} J. Torrico, M. Rojas, S. M. de Souza, Onofre Rojas, N. S. Ananikian, EPL 108 (2014) 50007
\bibitem{Ananikian diamond chain7} V. S. Abgaryan, N. S. Ananikian, L. N. Ananikyan, V. Hovhannisyan, Solid State Communications 224 (2015) 15-20
\bibitem{Ananikian diamond chain8} V. S. Abgaryan, N. S. Ananikian, L. N. Ananikyan, V. Hovhannisyan, Solid State Communications 203 (2015) 5-9
\bibitem{Afsaneh1} Marzieh Asoudeh, Vahid Karimipour, Afsaneh Sadrolashrafi, Phys. Rev. B 76, 064433 (2007)
\bibitem{Afsaneh2} Marzieh Asoudeh, Vahid Karimipour, Afsaneh Sadrolashrafi, Phys. Rev. B 75, 224427 (2007)
\bibitem{Afsaneh3} Marzieh Asoudeh, Vahid Karimipour, Afsaneh Sadrolashrafi, Phys. Rev. A 76, 012320 (2007)
\bibitem{Magnetic susceptibility - entanglement witness} M. Wie\'{s}niak, V. Vedral, C. Brukner,  New J. Phys., 7 (2005) 258
\bibitem{Entanglement and correlation} Ulrich Glaser, Helmut Büttner, Holger Fehske, Phys. Rev. A 68, 032318 - Published 30 September 2003
\bibitem{Dynamics of entanglement in one-dimensional spin systems} Luigi Amico, Andreas Osterloh, Francesco Plastina, Rosario Fazio, G. Massimo Palma, Phys. Rev. A 69, 022304 - Published 13 February 2004
\bibitem{ueda} Kazuo Ueda, Shin Miyahara, J. Phys.: Condens. Matter 11 (1999) L175?L178
\bibitem{koga} Akihisa Koga, Journal of the Physical Society of Japan, Vol. 69, No. 11, November, 2000, pp. 3509-3512
\bibitem{Baxter} R. J. Baxter, Academic, New York, 1982
\bibitem{ferromagnetic-ferromagnetic-antiferromagnetic spin chain} K. Okamoto, Solid State Commun. 98 (1996) 245
\bibitem{Haldane Gap for Half-Integer Spins} M. Oshikawa, M. Yamanaka and I. Affleck, Phys. Rev. Lett.78 (1997)1984
\bibitem{Breakdown of the magnetization plateau} A. Kitazawa and K.Okamoto, J. Phys. Condens. Matter 11 (1999) 9765
\bibitem{trimerized XXZ spin chain} K. Okamoto and A. Kitazawa, J. Phys. A Math. Gen. 32 (1999) 4601
\bibitem{polymerized spin chains} D. C. Cabra and M. D. Grynberg, Phys. Rev. B 59 (1999) 119
\bibitem{Antiferromagnetic Zigzag Spin Chain in the Strongly Frustrated Region} K. Okunishi and T. Tonegawa, J. Phys. Soc. Jpn. 72 (2003) 479
\bibitem{Fractional Sz excitation in Ising-like zigzag XXZ chain} K. Okunishi and T. Tonegawa, Phys. Rev. B 68 (2003) 224422
\bibitem{anisotropic s=1/2 antiferromagnetic chain} T. Tonegawa, K.Okamoto, K.Okunishi, K.Nomura and M.Kaburagi, Physica B 346-347 (2004), 50
\bibitem{solid 3He system on a triangular lattice} T. A. Arakelyan, V. R. Ohanyan, L. N. Ananikian, N. S. Ananikian, M. Roger, Phys. Rev. B 67, 024424 (2003) 
\bibitem{ZIGZAG LADDER} V. V. Hovhannisyan, L. N. Ananikyan, N. S. Ananikian, Int. J. of Mod. Phys. B 21, 3567 (2007)
 \bibitem{zigzag ladder2} V. V. Hovhannisyan, N. S. Ananikian, Phys. Lett. A 372, 3363 (2008)
\bibitem{ferromagnetic-ferromagnetic-antiferromagnetic Ising chain} V. R. Ohanyan, N. S. Ananikian, Phys. Lett. A 307 76 (2003)
\bibitem{Mag and Lyapun on a kagome chain} N. Ananikian, L. Ananikyan, R. Artuso, H. Lazaryan, Phys. Lett. A 374, 4084 (2010)
\bibitem{spin-1/2 Heisenberg diamond chains} B. Gu and G. Su, Phys. Rev. B 75 (2007) 174437
\bibitem{dimer-monomer} J. Kang, C. Lee, R. K. Kremer, M-H. Whangbo, J. Phys.: Condens. Matter 21 (2009) 392201
\bibitem{Wootters} W. K. Wootters, Phys. Rev. Lett. 80, 2245 (1998)
\bibitem{Hill} S. Hill and W.K. Wootters, Phys. Rev. Lett. 78, 5022 (1997)
\bibitem{paramagnetism of copper acetate} B. Bleaney, F. R. S. and K. D. Bowers, Proc. R. Soc. London, Ser. A, 214 (1952) 415
\bibitem{Quantum entanglement in Nitrosyl iron complexes} S. M. Aldoshin and E. B. Feldman, M. A. Yurishchev, Journal of experimental and theoretical physics, 107 (2008) 804
\bibitem{Exp determination of thermal ent in spin clusters mag suscep} A. M. Souza, M. S. Reis, D. O. Soares-Pinto, I. S. Oliveira, R. S. Sarthour, Phys. Rev. B, 77 (2008) 104402
\bibitem{trimer spin-1/2 Heisenberg chains} O. M. D. Cima, D. H. T. Franco, S. L. L. da Silva, Quantum Stud.: Math. Found. (2016) 3: 57
\bibitem{Horodecki} M. Horodecki, P. Horodecki, R. Horodecki, Phys. Lett. A, 223 (1996) 1
\bibitem{Horodecki2} M. Lewenstein. B. Kraus, J. I. Cirac, P. Horodecki, Phys. Rev. A, 62 (2000) 052310
\bibitem{entanglement in bulk properties of solids} C. Brukner, V. Vedral, A. Zeilinger, Phys. Rev. A, 73 (2006) 012110
\bibitem{Thermal ent in the nanotubular system} T. V\'{e}rtesi, E. Bene, Phys. Rev. B, 73 (2006) 134404
\bibitem{Exp observ of ent in low dimensional spin sys} T. G. Rappoport, L. Ghivelder, J. C. Fernandes, R. B. Guimar\~{a}es, M. A. Continentino, Phys. Rev. B, 75 (2007) 054422
\bibitem{Exp detection of therm ent in a molecular chain} T. Chakraborty, T. K. Sen, H. Singh, D. Das, S. K. Mandal, C. Mitra, J. Appl. Phys. 114 (2013) 144904
\end{thebibliography}

\end{document}